\documentclass[11pt,preprint2,11pt]{aastex} 
\usepackage[margin=0.8in]{geometry}

\usepackage{amsmath,amstext}
\usepackage[T1]{fontenc}
\usepackage{graphicx}
\usepackage{amsmath,amssymb}
\usepackage{color}
\usepackage{bm}
\usepackage[breaklinks,colorlinks,citecolor=blue,linkcolor=blue]{hyperref} 
\usepackage[all]{hypcap} 
\usepackage[normalem]{ulem}
\usepackage{soul}
\usepackage{multirow}
\usepackage{multicol}

\bibpunct{(}{)}{;}{}{}{,}

\def\mc{\multicolumn}
\def\mr{\multirow}

\def\msun{M_{\odot}}
\def\lsun{L_{\odot}}

\def\funit{\rm erg\,cm^{-2}\,s^{-1}}
\def\rsun{R_{\odot}}
\def\mach{\mathcal{M}}
\def\msunyr{\,\msun\,\rm yr^{-1}}
\def\rstar{R_{\star}}
\def\mstar{M_{\star}}

\def\gmin{\gamma_{\rm min}}
\def\gmax{\gamma_{\rm max}}
\def\gb{\gamma_{\rm b}}
\def\dmw{\dot{M}_{\rm w}}
\def\vw{v_{\rm w}}

\def\tw{T_{\rm w}}
\def\rhow{\rho_{\rm w}}
\def\vesc{v_{\rm esc}}
\def\ent{\epsilon_{\rm nt}}
\def\bp{B_{\rm p}}
\def\rso{R_{\rm so}}
\def\rp{R_{\rm p}}
\def\rorb{R_{\rm orb}}
\def\rjup{R_{\rm jup}}

\def\tacc{t_{\rm acc}}
\def\rl{R_{\rm L}}
\def\xacc{\xi_{\rm acc}}
\def\me{m_{\rm e}}
\def\tcool{t_{\rm cool}}
\def\tdyn{t_{\rm dyn}}
\def\ustar{U_{\star}}
\def\tstar{T_{\star}}
\def\nustar{\nu_{\star}}
\def\nuic{\nu_{\rm IC}}
\def\ub{U_{\rm B}}
\def\sigt{\sigma_{\rm T}}
\def\lstar{L_{\star}}
\def\ustar{U_{\star}}
\def\lkin{L_{\rm kin}}
\def\lnt{L_{\rm nt}}
\def\re{r_{\rm e}}
\def\vsyn{\nu_{\rm syn}}
\def\psyn{P_{\rm syn}}
\def\pic{P_{\rm IC}}
\def\tstar{T_{\star}}
\def\cs{c_{\rm s}}

\def\vk{v_{\rm k}}
\def\pw{P_{\rm w}}
\def\rorba{R_{\rm orb,-2}}
\def\lstara{L_{\star,\odot}}
\def\bjup{B_{\rm Jup}}

\title{Non-thermal emission from the interaction of magnetized exoplanets with the wind of their host star}
\author{Xiawei Wang and Abraham Loeb}
\affil{Department of Astronomy, Harvard University, 60 Garden Street, Cambridge, MA 02138, USA}%

\begin{abstract}
We study the non-thermal emission from the interaction between magnetized Jupiter-like exoplanets and the wind from their host star.
The supersonic motion of planets through the wind forms a bow shock that accelerates electrons which produces non-thermal radiation across a broad wavelength range.
We discuss three wind mass loss rates: $\dmw\sim10^{-14}$, $10^{-9}$, $10^{-6}\,\msunyr$ corresponding to solar-type, T Tauri and massive O/B type stars, respectively.
We find that the expected radio synchrotron emission from a Jupiter-like planet is detectable by the \emph{Jansky Very Large Array} and the \emph{Square Kilometer Array} at $\sim1-10$ GHz out to a distance $\sim 100$ pc, whereas the infrared emission is detectable by the \emph{James Webb Space Telescope} out to a similar distance.
Inverse Compton scattering of the stellar radiation results in X-ray emission detectable by \emph{Chandra} X-ray Observatory out to $\sim 150$ pc.
Finally, we apply our model to the upper limit constraints on V380 Tau, the first star-hot Jupiter system observed in radio wavelength.
Our bow shock model provides constraints on the magnetic field, the interplanetary medium and the non-thermal emission efficiency in V380 Tau.
\end{abstract}

\begin{document}

\section{Introduction}
\label{sec:sec1}
Thousands of exoplanet systems have been identified over the past few decades \citep{winn2015}.
The majority of the currently known population was indirectly discovered via searches for the impact of the exoplanet on its host star.

In analogy with the solar system, exoplanets might possess intrinsic magnetic fields and generate non-thermal radio emission \citep{garraffo2016}.
These magnetized exoplanets can be probed in radio observations since they produce more radio emission than the host star.
A number of observations support the existence of magnetic fields in exoplanets.
For instance, the near-UV spectroscopic transit of a giant planet WASP-12b shows an early ingress compared to its optical transit and an excess absorption during the transit \citep{haswell2012}; such a signature has been explained by absorption in a bow shock surrounding the planetary magnetosphere \citep{llama2011}. 
Another clue is provided by the modulations of chromoshperic spectral lines in phase with the orbital period, indicating induced activity on the stellar surface due to magnetic interactions between star and planet \citep{shkolnik2008}.
The magnetized planets in the solar system emit low-frequency radio waves from their auroral regions via the cyclotron maser instability (CMI) \citep{treumann2006}.
This emission is observed to be highly circularly (or elliptically) polarized and variable on a time scale from seconds to days  \citep{treumann2006, zarka2007}.
Magnetized exoplanets are expected to produce radio emission via a similar mechanism.
The power of this emission can be estimated by the empirical relation known as the radiometric Bode's law, which relates the incident energy flux of the stellar wind to the radio intensity of a planet, as inferred from observations of magnetized planets in the solar system \citep{zarka2001}.
This method was applied to hot-Jupiters but no detection has been reported as of yet \citep{jardine2008}.

A number of theoretical studies computed the expected exoplanetary radio emission by applying the radiometric Bode's law.
They found that the power of the radio emission depends on the planetary magnetic field and the kinetic energy flux of the stellar wind or coronal mass ejections \citep{griesmeier2011}.
\citet{lazio2004} predicted that planets on tight orbits at distances of a few pc might produce mJy level emission at $\sim 10-1000$ MHz frequencies.
It has been suggested that stars with winds carrying a larger mass loss rate and velocities than the Sun are ideal targets for radio observations \citep{stevens2005}, highlighting close-in hot Jupiters around pre-main-sequence and post-main-sequence stars for radio selection \citep{vidotto2017}.
Many observational campaigns have made effort to detect radio emission from exoplanets.
Some of them targeted nearby hot Jupiters (e.g. \citet{bastian2000, ogorman2018}), while others search for radio emission at locations of known exoplanets from low-frequency sky surveys (e.g. \citet{lazio2004, murphy2015}).

Planetary emission can be used to discover new planets or set constraints on the properties of the interplanetary medium around stars \citep{wood2005}.
The interaction between exoplanets and stellar winds leads to distinct observational signatures, such as stellar activity enhancement \citep{shkolnik2005}, cometary tail structures \citep{rappaport2012} and charge transfer between wind protons and  neutral hydrogen atoms \citep{kislyakova2014}.
These signatures provide constraints on the mass loss rate and speed of the stellar wind as well as the planetary magnetic field.
The formation of a bow shock from the interaction between stellar wind and exoplanetary magnetic field has been considered (e.g. \citet{zarka2007, vidotto2015}). 
However, previous discussions were limited to low-frequency radio emission from CMI, with no detailed calculation of the non-thermal emission produced by relativistic particles accelerated by the bow shock.

Here, we compute the non-thermal spectrum as a novel observational signature of exoplanets as they travel in the wind of their host star.
The supersonic motion of a planet can produce multi-wavelength emission detectable at a distance of up to hundreds of pc with current and upcoming instrumentation.
Aside from revealing new planets, any detection of such an emission can be used to set constraints on the properties of the interplanetary medium, wind mass loss rate and planetary magnetic field.

Our discussion is organized as follows.
In \S~\ref{sec:sec2}, we characterized the properties of the planetary bow shocks.
In \S~\ref{sec:sec3}, we compute the resulting non-thermal synchrotron and inverse Compton emission.
In \S~\ref{sec:sec4}, we apply our model to the solar system and the V380 Tau system.
Finally, in \S~\ref{sec:sec5}, we summarize our results and discuss observational implications.

\section{Planetary bow shock}
\label{sec:sec2}
As an exoplanet orbits around its host star, it interacts with the wind outflowing from the star.
For simplicity, we assume that the wind speed, $\vw\sim\vesc$, where $\vesc\sim(2G\mstar/\rstar)^{1/2}$ is the escape velocity from the star, $G$ is the Newton's constant and $\mstar$ and $\rstar$ are the mass and radius of the star, respectively.
The orbits of planets at small separation from their host star are often circularized by tidal dissipation, and their Keplerian orbital velocity is given by, $\vk= (G\mstar/\rorb)^{1/2}$, where $\rorb$ is the orbital radius of the planet.
Thus, the effective velocity of the planet relative to the interplanetary plasma is of order $\Delta v\sim(\vw^2+\vk^2)^{1/2}$ \citep{lynch2018}.
For simplicity, we adopt an isothermal profile for the stellar wind, $\rhow=\dmw/(4\pi\vw\rorb^2)$, where $\dmw$ is the stellar mass loss rate \citep{see2014}.
The magnetic field of exoplanets shields the stellar wind and deflects the interplanetary particles from reaching the planetary atmosphere.
Assuming a dipolar planetary magnetic field, we obtain the magnetic field at the stand-off radius, $\bp=B_0(\rp/\rso)^3$, where $B_0$ is the magnetic field at the equator on the planet's surface ($\sim$ half of the intensity at the magnetic pole) and $\rp$ is the planet's radius. 

The stand-off radius, $\rso$, is estimated by balancing the total pressure of the stellar wind and the planet's magnetic pressure:
\begin{equation}
p_{\rm w}\approx\frac{1}{2}\rhow\Delta v^2=\frac{\bp^2}{8\pi}\;.
\end{equation}
The thermal pressure of the wind is assumed to be negligible compared with its ram pressure \citep{vidotto2015}.
Therefore, the Mach number of the bow shock is given by, $\mach=\Delta v/\cs$, where $\cs=(\Gamma \pw/\rhow)^{1/2}$ is the  sound speed, with $\Gamma\sim1$ for an isothermal gas, and $\pw$ is the wind thermal pressure.
For $\mstar\sim\msun$, $\rstar\sim\rsun$ and wind temperature $\tw\sim10^6$ K, the mach number $\mach\sim 10$, where $\rsun$ is the solar radius, consistent with numerical simulations \citep{vidotto2015}.
Therefore, the orbits of close-in hot Jupiters are supersonic, leading to a bow shock with a Mach cone of opening angle $\sim1/\mach$ in the direction of planet's relative motion, that accelerates interplanetary electrons to relativistic energies, producing non-thermal emission.
\section{Non-thermal emission}
\label{sec:sec3}
Next, we calculate the non-thermal emission from the bow shock as the planet plunges through the stellar wind with $\mach\gg1$.
\subsection{Synchrotron emission}
In analogy with the collisionless shocks around supernova remnants \citep{helder2012}, the free electrons in the interplanetary medium are expected to be accelerated to relativistic energies via the Fermi acceleration mechanism. 
Their energy distribution can be described by a broken power-law:
\begin{equation}
N(\gamma)d\gamma=N_0 \gamma^{-p}\left(1+\frac{\gamma}{\gb}\right)^{-1} \quad (\gmin\le\gamma\le\gmax)\;,	
\end{equation}
where $N_0$ and $p$ are the normalization factor and power-law index of the electron density distribution, with $\gb$, $\gmin$ and $\gmax$ being the break, minimum and maximum Lorentz factor, respectively.
The electron acceleration timescale is given by $\tacc=\xacc\rl c/\vw^2$, where $\xacc$ is a dimensionless constant of unity \citep{blandford1987}, $\rl=\gamma\me c^2/e\bp$ is the Larmor radius, and $\me$ is the electron mass.
The maximum Lorentz factor, $\gmax$, is obtained by equating $\tacc$ to the minimum between the dynamical timescale, $\tdyn\sim\rso/\vw$, and the cooling timescale, $\tcool=3\me c/4 (\ub+\ustar)\sigt\gamma$.
Here $\ub=\bp^2/8\pi$ and $\ustar=\lstar/(4\pi\rso^2 c)$ are the energy densities of the magnetic field and host star, respectively, and $\sigt$ is the Thomson cross-section.
For typical parameters, we find that $\tacc>>\tdyn$, and so $\gmax$ is mainly constrained by $\tdyn$.
The break Lorentz factor, $\gb$, can be obtained by equating $\tdyn$ and $\tcool$, which yields $\gb=3\me c\vw/4\sigt\rso(\ub+\ustar)$.
We adopt $\gmin\sim1$ in the calculation.
The power-law index of accelerated electrons, $p$, is related to the Mach number of the shock, $\mach$, through \citep{drury1983, gargate2012}:
\begin{equation}
p=\frac{r+2}{r-1}
\end{equation}
where $r$ is the shock compression ratio, derived from the shock jump condition:
\begin{equation}
r=\frac{(\Gamma+1)\mach^2}{(\Gamma-1)\mach^2+2}
\end{equation}
$p\sim2-2.2$ is inferred from numerical simulations of strong shocks \citep{gargate2012}. 
Numerical simulation and observations of supernova-driven shock suggests that $p\sim 2.1-2.5$ \citep{helder2012, caprioli2014}.
Here, we consider $p$ as a free parameter in the calculation.
We assume that a fraction of the kinetic energy of the stellar wind is converted to the total non-thermal luminosity:
\begin{equation}
\lnt=\ent\lkin\approx\frac{1}{2}\ent\rhow\Delta v^3(\pi\rso^2)\;,
\end{equation}
where $\ent$ is the fraction of electrons accelerated to produce non-thermal radiation which we leave as a free parameter in our model.
For supernova remnants, $\ent\sim5\%$ \citep{helder2012}.

Next, we compute the synchrotron emission following the standard formula from \citet{rybicki1979}. 
The emission and absorption coefficients are given by:
\begin{equation}
j_{\nu}^{\rm syn}=c_{1}B\int^{\gmax}_{\gmin} F(\frac{\nu}{c_1 B \gamma^2})
N(\gamma)\,d\gamma\; ,
\end{equation}
\begin{equation}
\alpha_{\nu}^{\rm syn}=-c_{2}B\frac{1}{\nu^{2}}\int^{\gmax}_{\gmin} \gamma^{2}
\frac{d}{d\gamma}\left[\frac{N(\gamma)}{\gamma^{2}}\right]
F(\frac{\nu}{c_1 B \gamma^2})\,d\gamma\; ,
\end{equation}
where 
$c_1=\sqrt{2}e^{3}/4\pi \me c^{2}$, $c_2=\sqrt{2}e^{3}/8\pi \me^{2}c^{2}$,
$F(x)\equiv x\int^{\infty}_{x} K_{5/3}(\xi)\,d\xi$ and
$K_{5/3}(x)$ is the modified Bessel function of $5/3$ order.
The synchrotron emission peaks at a frequency of $\vsyn=4.2\times10^{14} B_1\gamma_4^2$ Hz, where $B_1=(\bp/1\,\rm G)$ and $\gamma_4=(\gamma/10^4)$.
The specific intensity of synchrotron emission can be obtained by the radiative transfer equation \citep{rybicki1979}:
\begin{equation}
I_{\nu}=\frac{j_{\nu}^{\rm syn}}{\alpha_{\nu}^{\rm syn}}\left(1-e^{-\tau_{\nu}}\right)\;,
\end{equation}
where $\tau_{\nu}$ is the optical depth.
The solid lines in Fig. \ref{fig:fig1} show the synchrotron emission for three cases of $\dmw$, which corresponds to solar-type stars ($\dmw\sim10^{-14}\msunyr$), T Tauri type stars with intermediate mass loss ($\dmw\sim10^{-9}\msunyr$) and massive O/B type stars ($\dmw\sim 10^{-6}\msunyr$).
We apply our model to the non-thermal emission produced by the bow shock from Jupiter in the solar system to constrain our free parameters.
For solar wind of mass loss rate $\dmw=3\times10^{-14}\,\msunyr$, $\rorb\sim5$ AU, Jovian polar magnetic field $\bjup\sim7$ G, we find that $\rso\sim 40\,\rjup$, $\gmax\sim 100$, consistent with the estimate from \citet{depater2003}.
The observed radio flux at 1.4 GHz from Jupiter is $\sim 4-5$ Jy \citep{depater2003, zarka2007}, setting an upper limit on $\ent\lesssim0.3$, consistent with the value of $\ent\sim5\%$ inferred in supernova remnants \citep{helder2012}.

We find that for massive stars, radio synchrotron emission below $\sim 10$ GHz is self-absorbed, and emission at $\gtrsim 10$ GHz can be detected at a distance of $\lesssim 300$ pc (see Table \ref{tab:tab1} for details).
For intermediate mass stars, synchrotron self-absorption takes place at $\sim 1$ GHz and radio emission at $\gtrsim 1$ GHz is observable out to $\sim200$ pc.
For solar-type stars, GHz emission is not affected by self-absorption. 
However, the low kinetic luminosity of the wind leads to weak non-thermal intensity and the detectability is limited to $\lesssim 100$ pc.

\subsection{Inverse Compton scattering}
Inverse-Compton (IC) scattering of low-energy photons by the same electrons responsible for the synchrotron emission could produce high-energy radiation.
The soft photons are provided by the photosphere of the host star as well as the synchrotron photons.
The energy density of the stellar radiation field is $\ustar=\lstar/(4\pi\rorb^2 c)$.
In comparison, the magnetic field has an energy density of $\ub=\bp^2/8\pi$.
The ratio between synchrotron and IC power is equivalent to the ratio between magnetic and stellar radiation energy density, i.e. $\psyn/\pic=\ub/\ustar\approx 0.01\, B_1^2\rorba^2\lstara^{-1}$, where $\rorba=(\rorb/10^{-2}\,\rm AU)$, $\lstara = (\lstar/\lsun)$, $\lsun$ is the solar luminosity.
Thus, we expect significant IC emission from close-in exoplanet systems, such as hot Jupiters.
For simplicity, we approximate the spectral energy distribution of the stellar emission as a blackbody spectrum of temperature $\tstar$.
The specific intensity of a stellar spectrum can be written as:
\begin{equation}
I_{\nu}=\frac{\lstar}{4\pi\rorb^2 c}f_{\nu}(\tstar)\;,
\end{equation}
where $f_{\nu}$ is the normalized Planck function.
The stellar luminosity-mass relationship can be expressed as $\lstar\propto\mstar^{\alpha}$, where $\alpha\sim 2.3$, $4.0$, $3.5$ and $1.0$ for $\mstar<0.43\msun$, $0.43<\mstar/\msun<2.0$, $2.0<\mstar/\msun<20$ and $\msun>50\msun$, respectively \citep{malkov2007}.

The IC scattering of stellar photons peaks at a frequency of $\nuic\approx\gamma^2\nustar=5.9\times10^{21}\,\gamma_{4}^2T_{\star,3}$ Hz, where $\gamma_4=(\gamma/10^4)$ and $T_{\star,3}=(T_{\star}/10^3\,\rm K)$.
The differential production rate of high-energy photons with energy $\epsilon\me c^2$ is given by \citep{coppi1990}:
\begin{equation}
Q(\epsilon)=\int\,d\epsilon_0\,n(\epsilon_0)\int\,d\gamma N(\gamma)\,K(\epsilon,\gamma,\epsilon_0)\;,
\end{equation}
where $\epsilon_0\me c^2$ is the soft photon energy, $\gamma\me c^2$ is the electron energy and $n(\epsilon_0)$ is the number density of soft photons.
$K(\epsilon,\gamma,\epsilon_0)$ is the Compton kernel, expressed as:
\begin{equation}
\begin{split}
K(\epsilon,\gamma,\epsilon_0)=\frac{2\pi\re^2 c}{\gamma^2\epsilon_0}
[&2\kappa\ln\kappa+(1+2\kappa)(1-\kappa)\\
&+\frac{(4\epsilon_0\gamma\kappa)^2}{2(1+4\epsilon_0\gamma\kappa)}(1-\kappa)]\;,
\end{split}
\end{equation}
where $\kappa=\epsilon/[4\epsilon_0\gamma(\gamma-\epsilon)]$ and $\re=e^2/\me c^2$ is the classical electron radius.
The emissivity of IC scattering is given by:
\begin{equation}
j_{\nu}^{\rm IC}=\frac{h}{4\pi}\epsilon Q(\epsilon)\;,
\end{equation}
where $h$ is Planck's constant.

\begin{figure*}
\centering
  \begin{tabular}{c}
    \includegraphics[width=1.8\columnwidth]{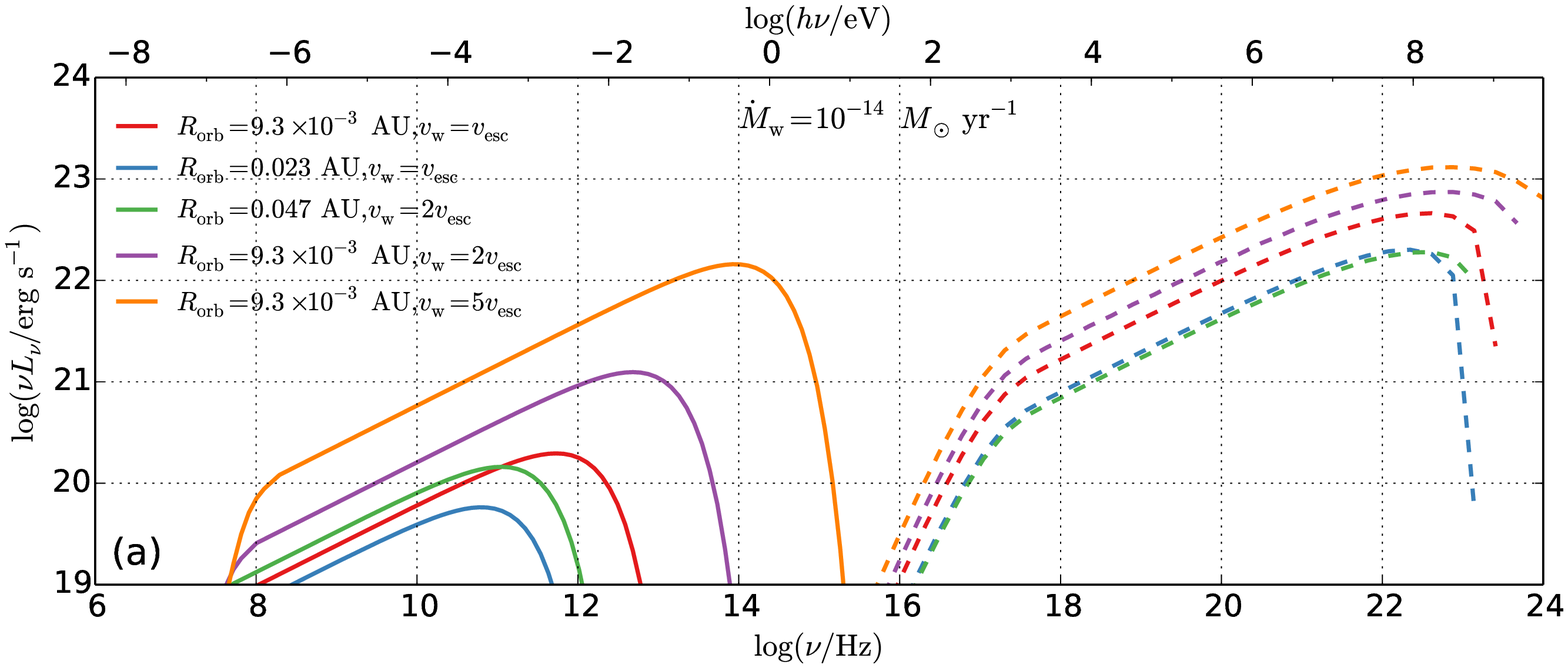} \\
    \includegraphics[width=1.8\columnwidth]{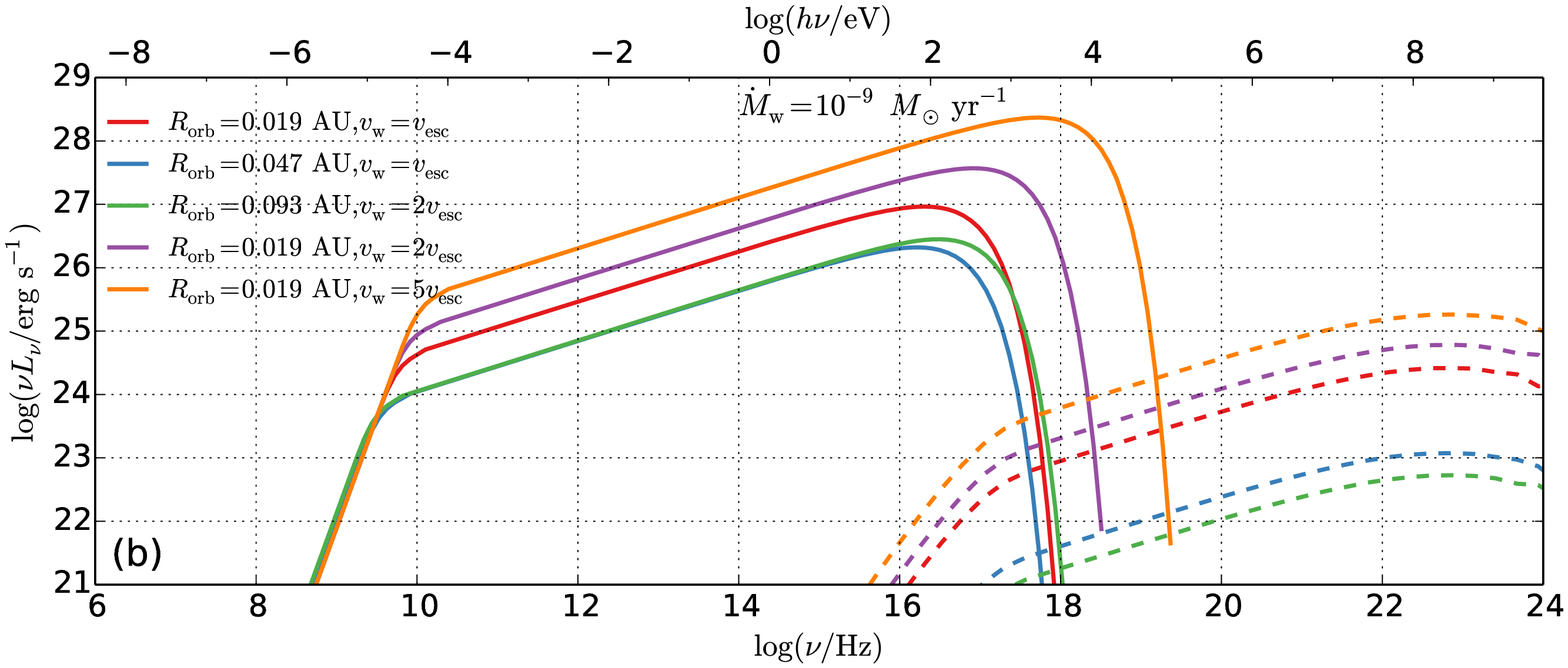} \\
    \includegraphics[width=1.8\columnwidth]{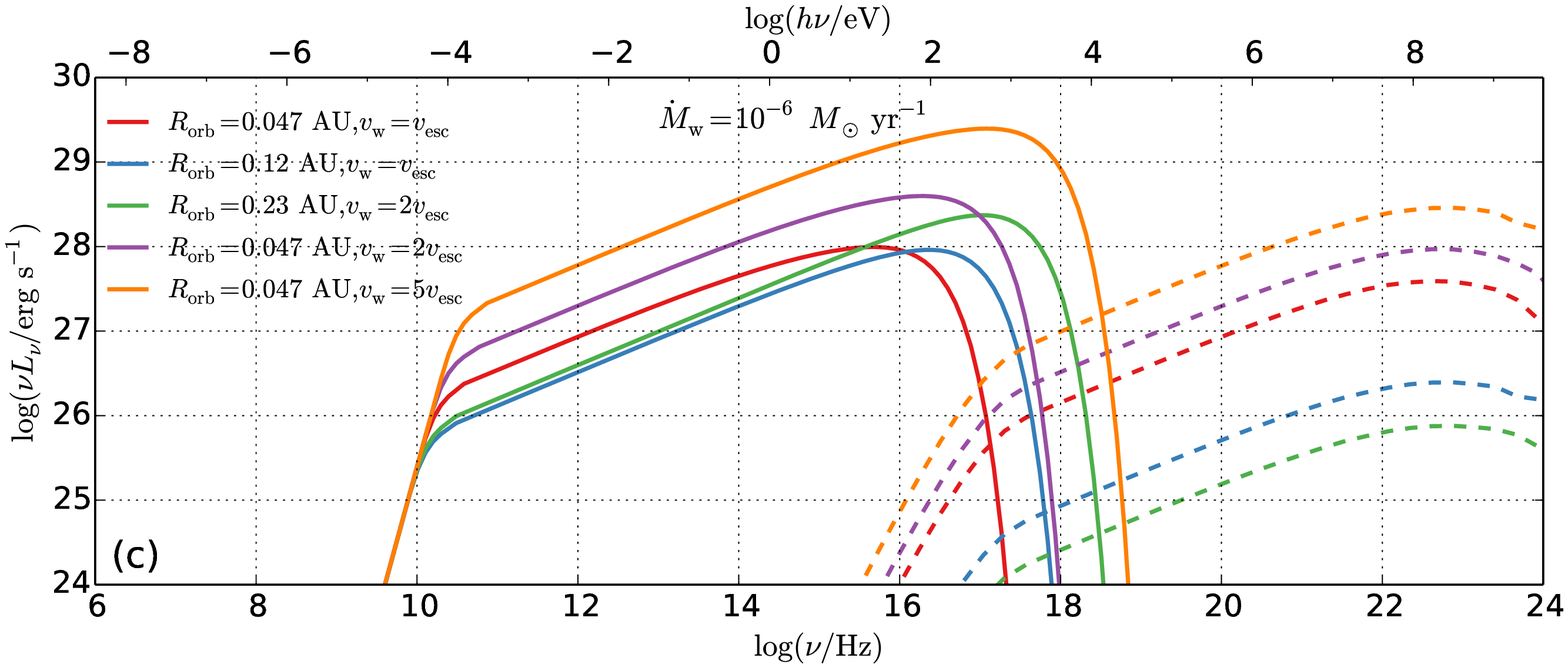} \\
  \end{tabular}
  \caption{\label{fig:fig1}
Non-thermal emission from stars with a characteristic wind mass loss rate of $\dmw\sim10^{-14}$, $10^{-9}$ and $10^{-6}\,\msunyr$, interacting with Jupiter-like planets, as shown in panel (a)-(c) respectively.
The solid and dashed lines corresponds to synchrotron emission and inverse Compton scattering off stellar photons, respectively.
The vertical axis shows the luminosity per $e$-folding in frequency and the horizontal axes show the frequency in Hz (bottom) or the equivalent photon energy in eV (top).
In panel (a), $\lstar\sim\lsun$ and $\rstar\sim\rsun$.
In panel (b), $\lstar\sim3\lsun$ and $\rstar\sim2\rsun$.
In panel (c), $\lstar\sim10^3\lsun$ and $\rstar\sim5\rsun$.
 }
\end{figure*}

The expected IC emission for three cases of $\dmw$ is shown by the dashed lines in Fig. \ref{fig:fig1}.
We find that the X-ray emission is detectable the \emph{Chandra} X-ray Observatory out to a distance of $\sim 200$ pc for massive stars (see Table \ref{tab:tab1} for details).
In particular, for solar-type stars, the total power in IC exceeds the synchrotron power.
Thus X-ray observations could detect close-in planets with $\rorb\lesssim 5\rstar$.

\subsection{Detectability}
Table \ref{tab:tab1} summarizes the detectability of our calculated non-thermal emission.
For solar-type stars ($\dmw=10^{-14}\,\msunyr$), the predicted radio fluxes are detectable out to $\lesssim 100$ pc with current and upcoming instrumentation. 
The emission at higher frequencies is too weak for detection.
For T Tauri stars ($\dmw=10^{-9}\,\msunyr$), we expect radio detection out to $\sim 150$ pc.
For massive O/B type stars, the emission is bright across all wavelengths from radio to X-rays, and observable out to a distance of $\sim 300$ pc.
Note that synchrotron self-absorption is significant at GHz for massive stars and the spectrum peaks at $\gtrsim8$ GHz (see Fig.\ref{fig:fig1}).
Thus, radio observation at higher frequencies is required to detect synchrotron emission from massive stars, in contrast to the CMI expected at low frequencies \citep{vidotto2010}.

\begin{table*}
\begin{center}
{\footnotesize
\begin{tabular}{ccccccccc}\hline\hline
&\mc{2}{c}{$\dmw=10^{-14}\,\msunyr$} & &\mc{2}{c}{$\dmw=10^{-9}\,\msunyr$} & & \mc{2}{c}{$\dmw=10^{-6}\,\msunyr$}\\ \cline{2-3}\cline{5-6}\cline{8-9}
\mr{2}{*}{Telescopes} & $F_{\nu}$   & \mr{2}{*}{detectability}    & & $F_{\nu}$  & \mr{2}{*}{detectability}  & & $F_{\nu}$  & \mr{2}{*}{detectability}
\\ 
   &\mc{1}{c}{($\mu$Jy)}  &  &  &\mc{1}{c}{($\mu$Jy)}   &    &  &\mc{1}{c}{($\mu$Jy)}  &
\\
\hline
JVLA		& 0.02; 0.004   &No; No &   &$0.2; 20$   &Marginal; Yes&   &$0.3; 380$     &Marginal; Yes\\
SKA        &  0.02; 0.004    &No &   &$0.2; 20$    &Yes  &   &$0.3; 380$   &Yes\\
ALMA     & 0.001    &No    &   & 4.0    &Marginal &   &150    &Yes \\
HST     & N/A    &No &   &0.038   &Marginal   &   &0.43    &Yes \\
JWST     & N/A    &No &   &0.038    &Marginal   &   &0.43    &Yes \\
\hline
&$\nu F_{\nu}$  &\mr{2}{*}{detectability}  &  &$\nu F_{\nu}$   &\mr{2}{*}{detectability}  &   &$\nu F_{\nu}$   &\mr{2}{*}{detecatbility}\\
  &\mc{1}{c}{($\funit$)}  &  &  &\mc{1}{c}{($\funit$)}  &  &  &\mc{1}{c}{($\funit$)}  &
\\
\cline{2-3}\cline{5-6}\cline{8-9}
XMM-Newton & $10^{-20}$     &No &   &$4\times10^{-15}$   &Yes    &   &$1.9\times10^{-13}$    &Yes\\
ATHENA  & $10^{-20}$   &No    &   &$4\times10^{-15}$    &Yes    &   &$1.9\times10^{-13}$    &Yes\\
Chandra    & $2\times10^{-20}$    &No  &   &$5\times10^{-15}$    &Yes     &   &$4\times10^{-14}$     &Yes \\
NuSTAR  & $2\times10^{-20}$     &No  &   &$5\times10^{-15}$   &Marginal   &   &$4\times10^{-14}$   &Yes \\
\hline
\end{tabular}
\caption{\label{tab:tab1}Detectability of non-thermal emission from exoplanet bow shock at a distance of 150 pc.}
\vskip 0.1cm
\parbox{2.0\columnwidth}
{
We choose the characteristic values described in the text as representative examples for the exoplanet systems at a distance of $\sim 150$ pc.
For radio frequencies, we provide fluxes at 1 GHz and 10 GHz, in units of mJy.
For X-ray observation, we present $\nu F_{\nu}$ in units of $\funit$.
The telescope detection limits are as follows:
\begin{enumerate}
\item 
\emph{Jansky Very Large Array (JVLA)}: $\sim 1\mu$Jy for 1 $\sigma$ detection and 12h integration time at most bands \citep{nrao2014}.
\item 
\textit{The Square Kilometer Array (SKA-MID)}: $\sim 0.7\mu$Jy RMS sensitivity for a 10h integration time \citep{prandoni2014}.
\item 
\textit{The Atacama Large Millimeter/submillimeter Array (ALMA)}:  At frequency $345$ GHz, the sensitivity $\sim 8.7\,\mu$Jy for 10h integration time is calculated by the  \textit{ALMA} Sensitivity Calculator (ASC) (https://almascience.eso.org/proposing/sensitivity-calculator).
\item
\textit{Hubble Space Telescope (HST)}: sensitivity $\sim 40-50$ nJy for the wavelength range of $0.6-1.5\,\mu$m for 10$\sigma$ detection and $10^4$ s integration time \citep{stsci2013}.
\item
\textit{The James Webb Space Telescope (JWST)}: sensitivity $\sim 10$ nJy for the wavelength range of $1-3\,\mu$m and $\sim 30$nJy for wavelengths $4-5\,\mu$m for 10$\sigma$ detection and $10^4$ s integration time \citep{stsci2013}.
\item
\emph{Chandra}: sensitivity of high resolution camera (HRC) $\sim 9\times 10^{-16} \funit$ covering energy range $0.08-10$ keV for 3$\sigma$ detection and $3\times 10^5$ s integration time \citep{chandra2014}.
\item
\textit{XMM-Newton}: $\sim 3.1\times10^{-16}\,\funit$ in the $0.5-2.0$ keV band \citep{hasinger2001}.
\item
\textit{Advanced Telescope for High Energy Astrophysics (ATHENA)}: $\sim 4\times 10^{-17}\,\funit$ in the $0.5-2$ keV band in a $10^6$s deep field \citep{barcons2012}.
\item
\textit{Nuclear Spectroscopic Telescope Array (NuStar)}: $\sim2\times10^{-15}\,\funit$ in $6-10$ keV band for 3$\sigma$ detection and $10^6$ s integration time \citep{harrison2013}.
\end{enumerate}
}}
\end{center}
\end{table*}
\section{Application to V380 Tau}
\label{sec:sec4}
\begin{figure*}[h!]
\centering
\includegraphics[angle=0,width=1.5\columnwidth]{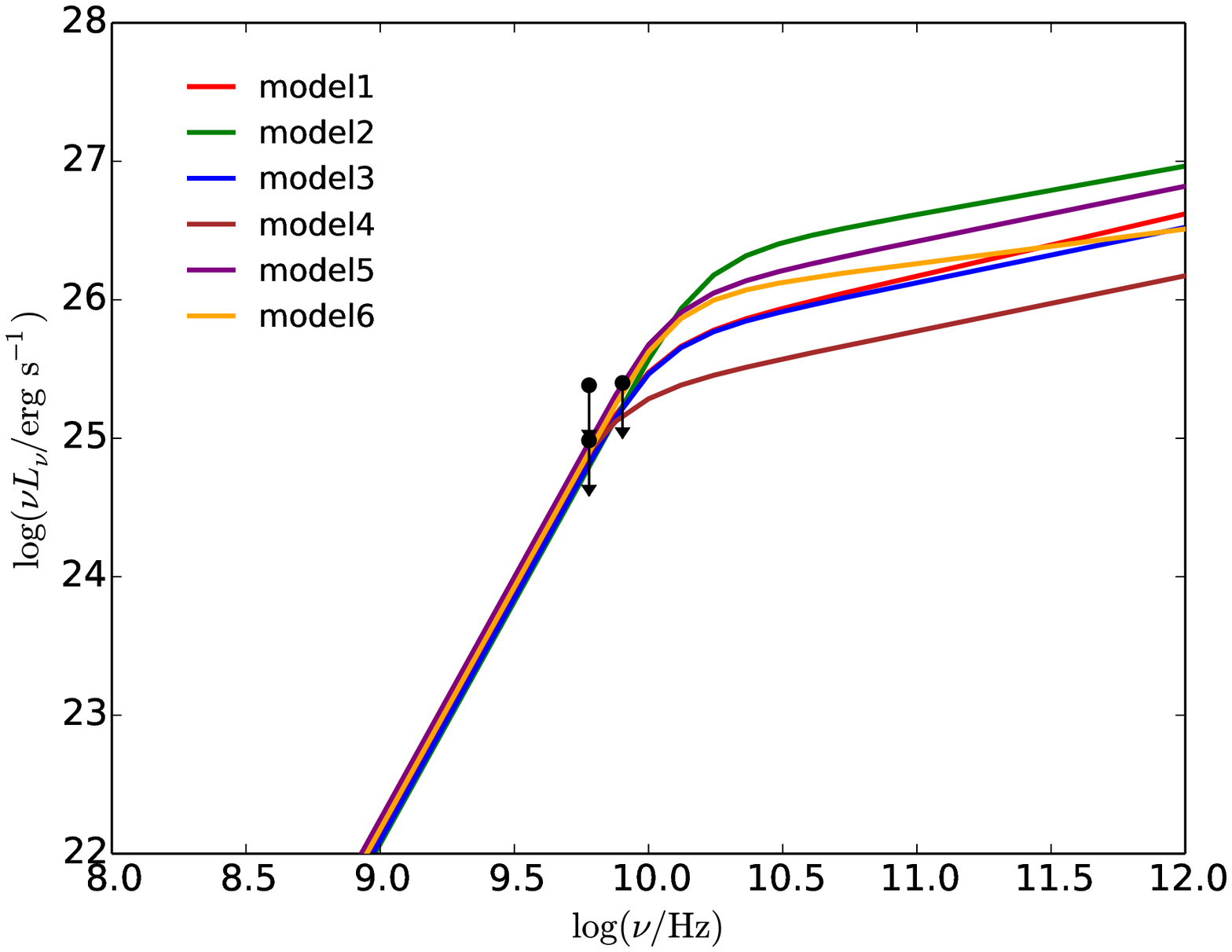}
\caption{\label{fig:fig2}
Estimated synchrotron emission from V380 Tau for the six models in Table \ref{table:tab2}.
We estimate the non-thermal synchrotron emission from the interaction between stellar wind and magnetosphere of the hot Jupiter, constrained by non-detection upper limits from \emph{VLA} and \emph{VLBA} observations.
}
\end{figure*}

We apply our model to V380 Tau, a non-accreting solar mass T-tauri star that hosts a hot Jupiter orbiting at a radius of 0.057 AU, located at a distance of 150 pc \citep{donati2016}.
\emph{Very Large Array} (VLA) observations at a frequency of 6 GHz reveal a flux density $919\pm26\,\mu$Jy, along with non-detections at two other epochs corresponding to limits of < 66 and < 150 $\mu$Jy \citep{bower2016}. 
In addition, \emph{Very Long Baseline Array} (VLBA) observations show one detection and one non-detection at comparable sensitivity, which indicates that the emission might be transient and possibly is non-thermal in origin \citep{bower2016}.
In Fig. \ref{fig:fig2}, we fit the non-detection limit of V380 Tau system using the bow shock model with various combination of parameters as listed in Table \ref{table:tab2}.
We find that the synchrotron spectrum is steeper at $\nu\lesssim10$ GHz due to synchrotron self-absorption.
An X-ray counterpart of this emission from IC emission is expected, as shown in Fig. \ref{fig:fig1}.
Additionally, the predicted non-thermal synchrotron emission has a steeper spectrum than the CMI emission estimated from the radiometric Bode's law \citep{vidotto2017}.
The non-thermal emission model can be applied to CI Tau b, which is around a star of comparable age to V380 Tau \citep{johns2016}.

\begin{deluxetable}{c c c c c c}
\tabletypesize{\scriptsize}
\tablecaption{Model parameters of synchrotron emission from V380 Tau}
\tablewidth{0pt}
\tablehead{
\colhead{Parameters}
&\colhead{$\dmw$ $(\msun\,\rm yr^{-1})$}
&\colhead{$\Delta v/\vesc$}
&\colhead{$\bp$ (G)}
&\colhead{$\ent$}
&\colhead{$p$}
}
\startdata
Model 1   & $5\times10^{-9}$   & $6.0$   & $2.0$   & $0.1$  & $2.1$
\\
Model 2   & $5\times10^{-9}$   & $6.0$   & $2.0$   & $0.1$  & $2.3$
\\
Model 3   & $1\times10^{-8}$   & $2.0$   & $2.0$   & $0.1$  & $2.2$
\\
Model 4   & $6\times10^{-10}$   & $3.0$   & $1.0$   & $0.25$  & $2.2$
\\
Model 5   & $1\times10^{-8}$   & $1.0$   & $1.0$   & $0.25$  & $2.2$
\\
Model 6   & $1\times10^{-10}$   & $7.0$   & $1.0$   & $0.5$  & $2.5$
\\
\enddata
\tablecomments{
$\dmw$: stellar wind mass loss rate;
$\Delta v/\vesc$: the ratio between exoplanet's relative speed to stellar wind and wind speed;
$\ent$: the fraction of wind kinetic luminosity converted to accelerate electrons to relativistic energies;
$p$: the power-law index of non-thermal electrons;
$\bp$: the magnetic field at the surface of the hot Jupiter.
}
\label{table:tab2}
\end{deluxetable}
\section{Discussion}
\label{sec:sec5}
In this \emph{Letter}, we studied the non-thermal emission produced by the supersonic motion of an exoplanet through the wind of its host stars.
This produces a unique fingerprint of the interaction between the planet's magnetosphere and the stellar wind, observable across a broad range of wavelengths from radio to X-rays.
In particular, we considered three characteristic cases of stellar wind mass loss rates, namely $\dmw=10^{-14},10^{-9},10^{-6}\,\msunyr$, corresponding to solar-type, T Tauri and massive O/B stars, respectively.
We have found that it is challenging to detect emission from solar-type stars farther than $\sim 100$ pc, but the detection of planets around massive stars is feasible out to a distance of $\sim 300$ pc.
For stars with intermediate mass loss rate, we find that X-ray frequencies allow the detection of exoplanets to a greater distance than their radio emission.
For stars with substantial mass loss, the search for radio emission should be restricted to higher frequencies $\gtrsim 10$ GHz as emission at lower frequencies is suppressed by synchrotron self-absorption. 
We note that the variability of the host star's magnetic field could mask the temporal variability from the bow shock \citep{llama2013}.

Past observations have searched for radio signatures of cyclotron emission from close-in exoplanets at low radio frequency using instruments such as the Low-Frequency Array (LOFAR) \citep{zarka2007}.
However, radio signatures of cyclotron emission from close-in exoplanets had not yet been detected due to instrumental sensitivity limitations at the $\sim 100$ MHz frequency range \citep{bastian2000}, though subtle hints of such emission had been claimed (e.g. \citet{ogorman2018}), and was postulated that the beaming of the emission could explain the non-detections \citep{lenc2018}.
Since only a small fraction of the exoplanets orbits is sampled by these observations, there could be an optimal orbital phase for the related radio detection \citep{lynch2018}. 
\citet{weber2017} showed that super-massive planets such as Tau Bootis b and CI Tau b \citep{johns2016} are highly favorable targets for CMI emission.
We find that the non-thermal signal is weakly subject to planet's mass, making it more promising for detection of less massive planets than the CMI emission. 
Another CMI source is the host star itself, which could contaminate the emission from planet \citep{llama2018, cotton2019}.
However, the associated frequencies are $\ll$ GHz, below the frequency of the non-thermal emission from planet-host star interaction.
In addition to low-frequency CMI searches, we propose to look for the non-thermal signature of these systems at higher frequencies.
Our calculations imply a new window for discovering exoplanet systems across a broad range of wavelengths from radio to X-rays. 
Detection of the emission signal from an exoplanet-wind interaction can provide constraints on the properties of stellar wind as well as the planet's magnetosphere.

\section*{Acknowledgements}
We thank an anonymous referee and John Forbes for insightful comments on the manuscript.
This work was supported in part by a grant from the Breakthrough Prize Foundation.
\end{document}